\documentclass[pra,superscriptaddress,twocolumn]{revtex4-1}
\usepackage[latin1]{inputenc}
\usepackage{amsmath}
\usepackage{amssymb}
\usepackage{dsfont}
\usepackage{color}
\usepackage{xcolor}
\usepackage{lineno}
\makeatletter
\usepackage{amssymb}
\usepackage{graphicx}
\usepackage{epstopdf}
\usepackage{epsfig}
\usepackage{subfigure}
\makeatother

\newcommand{\mje}[1]{{\color{black} #1}}

\usepackage{cancel,ifthen}
\usepackage[normalem]{ulem}

\newcommand{\comment}[2][NoInPuT]{\ifthenelse{\equal{#1}{NoInPuT}}{}{{\color{blue}\sout{#1}}}{\color{red} #2}}
\begin{document}

\title{Quantum droplets of quasi-one-dimensional dipolar Bose-Einstein condensates}
\author{M. J. Edmonds}
\affiliation{Department of Physics \& Research and Education Center for Natural Sciences, Keio University, Hiyoshi 4-1-1, Yokohama, Kanagawa 223-8521, Japan}
\author{T. Bland}
\affiliation{Joint Quantum Centre (JQC) Durham--Newcastle, School of Mathematics, Statistics and Physics, Newcastle University, Newcastle upon Tyne, NE1 7RU, United Kingdom}
\author{N. G. Parker}
\affiliation{Joint Quantum Centre (JQC) Durham--Newcastle, School of Mathematics, Statistics and Physics, Newcastle University, Newcastle upon Tyne, NE1 7RU, United Kingdom}

\date{\today{}}

\begin{abstract}\noindent
Ultracold dipolar droplets have been realized in a series of ground-breaking experiments, where the stability of the droplet state is attributed to beyond-mean-field effects in the form of the celebrated Lee-Huang-Yang (LHY) correction. We scrutinize the dipolar droplet states in a one-dimensional context using a combination of analytical and numerical approaches, and identify experimentally viable parameters for accessing our findings for future experiments. In particular we identify regimes of stability in the restricted geometry, finding multiple roton instabilities as well as regions supporting quasi-one-dimensional droplet states. By applying an interaction quench to the droplet, a modulational instability is induced and multiple droplets are produced, along with bright solitons and atomic radiation. We also assess the droplets robustness to collisions, revealing population transfer and droplet fission.       
\end{abstract}
\maketitle

\section{Introduction}

Bose-Einstein condensates possessing long-ranged, anisotropic dipole-dipole interactions have been realized in a series of ground-breaking experiments with highly magnetic atoms. The first generation of experiments achieved condensation of $^{52}$Cr \cite{griesmaier_2005,beaufils_2008}, $^{164}$Dy \cite{lu_2011,tang_2015} and $^{168}$Er \cite{aikawa_2012}. Recently, a second series of experiments has achieved a quantum analogue of the classical Rosensweig instability \cite{kadau_2016}, as well as the realization of droplet states \cite{barbut_2016,chomaz_2016} -- where the gas enters a high density phase whose stability has been attributed to the influence of quantum fluctuations \cite{baillie_2016,bisset_2016,baillie_2018,baillie_2017,wachtler_2016}. Dipolar condensates constitute weakly correlated systems, and can exhibit properties and behaviour similar to that of a liquid in the beyond-mean-field limit \cite{bulgac_2002,petrov_2016,edler_2017,zin_2018,oldziejewski_2019}.  Droplet states have also been realized in condensate mixtures \cite{semeghini_2018,cabrera_2018} -- supported by a balance between attractive {\it s}-wave interactions between the atoms and quantum fluctuations --  and photonic systems  \cite{wilson_2018} -- here the nonlinear medium provides a repulsive $d$-wave contribution that stabilizes the light beam.  

The quantum liquid Helium II is well-known to exhibit a roton minimum in its excitation spectrum; this is supported by the strong interatomic interactions and correlations \cite{leggett_book}, where roton excitations typically occur at wavelengths comparable to the average inter-particle separation, indicating that the superfluid is close to forming a crystalline structure \cite{schneider_1971,pitaevskii_1984}.  Although dipolar condensates are weakly correlated, the nonlocal character of the dipolar interaction supports roton-like excitations \cite{odell_2003,santos_2003}, which have now been experimentally realized in a gas of $^{166}$Er \cite{chomaz_2018,petter_2019}. Here the roton lengthscale is dictated by the geometry of the gas. A plethora of theoretical investigations have focussed on detailing the correspondence between rotons in the Helium II phase and weakly interacting dipolar gases.  Early work examined the possibility of roton excitations in a quasi-one-dimensional setting \cite{giovanazzi_2004}, as well as the manifestation of rotons in a rotating dipolar condensate \cite{lasinio_2013} and the identification of a roton mode in trapped dipolar condensates -- leading to the identification of `roton fingers' \cite{bisset_2013}, a discrete manifestation of the instability found in the homogeneous system.

The existence of the roton mode indicates the proximity of the quantum fluid to long-range crystalline order. If the quantum fluid can simultaneously support a superfluid state, the system is a candidate for a supersolid -- a phase of matter where these two forms of order coexist \cite{gross_1958,andreev_1969,chester_1970,leggett_1970} (see also the review of Leggett, Ref.~\cite{leggett_1998}). The possible unambiguous detection of a supersolid state in Helium has attracted long debate. Very recently, state-of-the-art experiments with dipolar condensates have reported supersolid phases in these systems in a quasi-one-dimensional setting \cite{bottcher_2019,tanzi_2019}. Complementary theoretical investigations in lower-dimensional dipolar gases have explored nonlinear wave structures  in the form of dark solitons \cite{bland_2015,pawlowski_2015,edmonds_2016,bland_2017} and bright solitons \cite{baizakov_2015,edmonds_2017,pedri_2005,tikhonenkov_2008,raghunandan_2015}. Related theoretical proposals have also examined the possibility of supersolids in binary condensates \cite{sachdeva_2020}, whose existence is supported by a stripe-like phase.

Within the weakly-interacting regime, dilute atomic condensates are well described by the celebrated Gross-Pitaevskii description \cite{pethick_2002,barenghi_2016}. A natural extension to this is the Bogoliubov-de Gennes formalism, which constitutes the linear response theory of the nonlinear Schr\"odinger equation, and gives insight into the behavior of the elementary excitations and collective modes of the condensate \cite{dalfovo_1999}. Early work demonstrated that this approach could also be applied to dipolar condensates \cite{ronen_2006}, revealing the effect of the dipolar interaction on the excitations. Such an approach is justified when the quantization of excitations is not required. For systems in the beyond mean-field regime it is an open question as to whether such an approach is justified. Nonetheless there are several works that use this approach to study the collective behavour of quantum droplets \cite{baillie_2017,tylutki_2020,hu_2020}. Knowledge of the excitation spectrum gives insight into many important properties of these systems -- in many physical effects the dimensionality of the system plays a key role in the dipolar condensates overall behavior \cite{baillie_2015}. Dipolar condensates with spin degrees of freedom have also been analyzed within the Bogoliubov-de Gennes framework, exploring the interplay of the excitations with the magnetic phases inherent to these systems \cite{huhtamaki_2011}. The anisotropy of the dipolar interaction leads to novel ground states, including a concave (red blood cell) shaped solution in flattened trapping potentials \cite{ronen_2007,wilson_2011,bisset_2012}, whose structure can be attributed to the excitation of a roton mode in this system \cite{ronen_2007,martin_2012}. Further work contrasted the dipolar Bogoliubov-de Gennes equations with a variational approach in the pancake geometry \cite{kreibich_2013}. The solutions to the Bogoliubov-de Gennes equations can also be used to study the depletion of the condensate, and in particular how the roton mode affects this important quantity \cite{blakie_2013}.

\begin{figure}[t]
\includegraphics[width=\columnwidth]{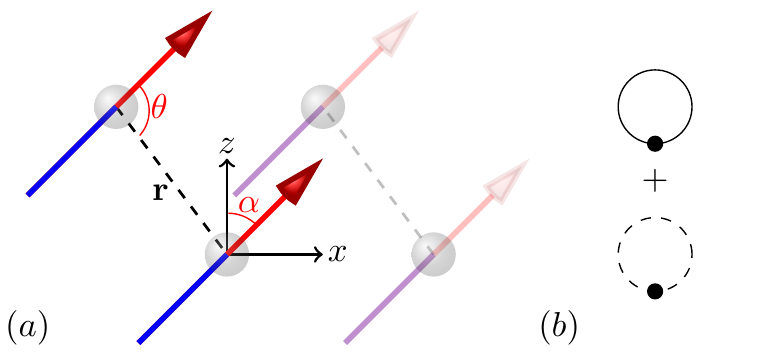}
\caption{\label{fig:dipole} Schematic representation of the atomic interactions (a). Each dipole (blue and red arrow) is separated by a distance ${\bf r}$, while $\theta$ defines the angle between the dipoles and the polarization direction, per Eq.~\eqref{eqn:dipdip}. The angle $\alpha$ defines the polarization direction of the dipoles in the $x$-$z$ plane. The isotropic contact interactions are represented by the gray spheres. (b) shows the two one-loop contributions to the ground state energy, Eq.~\eqref{eqn:eqf}.}
\end{figure}

Parallel to the ongoing experiments with atomic dipolar atoms, the realization of cold molecular gases has also revealed novel phenomenology. Different to their atomic counterparts, atomic molecules possess additional vibrational and rotational degrees of freedom, which complicates their manipulation and cooling \cite{bohn_2017}. Many different molecules have now been cooled, including RbCs \cite{molony_2014}, fermionic NaK \cite{woo_2015} and NaRb \cite{guo_2016}. The rapid development of this field has seen the achievement of the molecular rovibrational ground states of KRb \cite{ni_2008}, controlled quantum chemical reactions \cite{ospelkaus_2010}, and molecular collisions \cite{ni_2010}. The ability to control the dimensionality of ensembles of these molecules using optical lattices \cite{danzl_2010} has afforded a new route towards two-dimensional systems \cite{miranda_2011}, and quantum magnetism with molecular dipole-dipole interactions \cite{yan_2013}.

Understanding the role of dimensionality is important for quantum fluids in general, since different trapping configurations can alter the stability and character of the nonlinear solutions to these systems. Moreover, since fluctuations and interactions are enhanced in lower dimensional quantum systems, these regimes may provide greater insight into quantum fluctuations in general.  Several works have investigated droplets in a quasi-one-dimensional setting, including in binary mixtures \cite{astrakharchik_2018,mithun_2019}, and mixtures with coherent  \cite{chiquillo_2019} and spin-orbit \cite{tononi_2019} couplings. The related crossover from a bright soliton to a droplet state was also investigated experimentally \cite{cheiney_2018}. It is the aim of this work to understand the regimes of stability and the accompanying dynamics of dipolar droplets in the quasi-one-dimensional setting.

In this work we systematically investigate the solutions to the quasi-one-dimensional dipolar Gross-Pitaevskii model in the beyond-mean-field regime. We begin in Section \ref{sec:model} by studying the extended dipolar Gross-Pitaevskii model, from which we derive the Bogoliubov-de Gennes equations including the beyond-mean-field contribution, and use this to identify regimes of roton stability in the full parameter space of the model. Following this in Section \ref{sec:num} we solve numerically the extended Gross-Pitaevskii equation, examining the form of the solutions in the beyond-mean-field limit. The nature of the modulational instability is then scrutinised, as well as the behaviour of droplet collisions in this system. We conclude with a summary and outlook of our findings in Section \ref{sec:conc}. 

\section{\label{sec:model}Theoretical Model}
\subsection{Dipole-dipole Interactions}
We consider a gas of dipolar bosons of mass $m$ interacting through short-range $s$-wave and long-range dipole-dipole interactions. Then, the total atomic interaction potential has the form $U({\bf r})=g\delta({\bf r})+U_{\rm dd}({\bf r})$ with
\begin{equation}\label{eqn:dipdip}
U_{\rm dd}({\bf r})=\frac{C_{\rm dd}}{4\pi}\frac{1-3\cos^2\theta}{|{\bf r}|^3}
\end{equation}
where $g=4\pi\hbar^2a_s/m$ and $a_s$ defines the $s$-wave scattering length, while $C_{\rm dd}$ characterizes the strength of the dipole-dipole interaction, and $\theta$ defines the angle between the vector ${\bf r}$ joining two dipoles and the polarization direction of the dipoles. The atomic interactions are illustrated in Fig. 1(a).  The dipole polarization angle $\theta_{\rm m}=\cos^{-1}(1/\sqrt{3})$ defines the `magic' angle at which the dipolar interaction vanishes. If $C_{\rm dd}$ is positive and $\theta<\theta_{\rm m}$ the dipoles are orientated in an attractive head-to-tail configuration, while for $\theta>\theta_{\rm m}$ the dipoles lie side-by-side and are repulsive. The strength of the dipolar interaction is typically characterized in terms of the dimensionless parameter $\varepsilon_{\rm dd}=C_{\rm dd}/3g$. The `anti-dipole' regime, where $C_{\rm dd}$ is negative, has also been proposed in Ref. \cite{giovanazzi_2002} by performing a rapid rotation of the dipoles, such that the attractive and repulsive regimes are reversed. Recent experimental work \cite{tang_2018} indicated that such a scenario can be achieved; however, the condensate lifetime is hampered by a dynamical instability \cite{prasad_2019,baillie_2019}. Since we consider parameter regimes where the sign of $\varepsilon_{\rm dd}$ can be either positive or negative, we consider states with the possibility of differing signs of the parameters $C_{\rm dd}$ and $a_s$, which can be accessed by rapid rotation of the dipoles and with the powerful tool of optical Feshbach resonances. 
\subsection{Beyond-Mean-Field Dipolar Bogoliubov-de Gennes Equations}
The realization of stable droplet phases with highly magnetic $^{164}$Dy \cite{kadau_2016} has been attributed to {\it quantum fluctuations}. Theoretically, quantum fluctuations are formulated in terms of the Lee-Huang-Yang (LHY) correction \cite{lee_1957}, which within the local density approximation appears as a term proportional to a non-integer power of the atomic density in the appropriate generalized Gross-Pitaevskii equation.

To obtain the appropriate correction to the dipolar Gross-Pitaevskii equation, one begins with the many-body Hamiltonian for a gas of homogeneous dipoles and from this computes the ground state energy of the system. Such a situation was originally studied by the authors of Refs.~\cite{lima_2011}. The technical details and analysis for this are presented in Appendix \ref{app:qf}, the result of which gives
\begin{equation}\label{eqn:eqf}
\frac{E_{\rm QF}}{V}=\frac{64}{15}g\left(n_{0}^{\rm 3D}\right)^{2}\sqrt{\frac{n_{0}^{\rm 3D}a_{s}^{3}}{\pi}}\bigg(1+\frac{3}{2}\varepsilon_{\rm dd}^2\bigg).
\end{equation}
Equation \eqref{eqn:eqf} is independent of the dipole polarization angle $\alpha$, a result which will be utilized in this work to understand the interplay of the polarization angle in the beyond-mean-field regime. Here we adopt a \textit{quadratic} approximation for the beyond-mean-field contribution; this is primarily motivated to allow us to perform a thorough analytical analysis of the dipolar system in the beyond-mean-field limit. Further details supporting our choice of Eq.~\eqref{eqn:eqf} are given in Appendix \ref{app:qf}. 

To obtain a finite ground state energy for the homogeneous dipolar system (see Eq.~\eqref{eqn:eqf3d}), a renormalization procedure is required (see Appendix \ref{app:qf} for further details) in order to make the momentum integrals convergent, and hence physically meaningful. Then,  the two one-loop Feynman diagrams representing the renormalized ground state energy are shown in Fig.~\ref{fig:dipole} (b). The first diagram (solid lines) shows the real part of the diagonal propagator contribution from the fluctuations, while the second (dashed lines) shows the corresponding imaginary contribution \cite{andersen_2004}.

The condensate of $N$ atoms is parametrized by the wave function $\Psi({\bf r},t)$, normalized such that $\int \text{d}^3{\bf r}|\Psi({\bf r},t)|^2=N$.  Including the generalized dipolar LHY correction derived in Appendix \ref{app:qf}, the wave function is described by the generalized dipolar Gross-Pitaevskii equation, written as \cite{wachtler_2016},
\begin{equation}\label{eqn:dgpe}
i\hbar\frac{\partial\Psi}{\partial t}=\bigg[\frac{\hat{\bf p}^2}{2m}+\frac{1}{2}m\omega_\rho\rho^2+g|\Psi|^2+\Phi_{\rm dd}[\Psi]+\mu_{\rm QF}\bigg]\Psi.
\end{equation}
Here the mean field dipolar potential is defined as $\Phi_{\rm dd}[\Psi({\bf r},t)]=\int \text{d}^{3}{\bf r}'U_{\rm dd}({\bf r}-{\bf r}')|\Psi({\bf r}',t)|^2$, $\mu_{\rm QF}$ defines the density-dependent contribution from the quantum fluctuations treated in the local density approximation, and $\omega_{\rho}$ defines the harmonic trapping frequency in the radial $\rho^2=y^2+z^2$ direction. The beyond-mean-field chemical potential is calculated from Eq.~\eqref{eqn:eqf} using $\mu_{\rm QF}=\partial E_{\rm QF}/\partial N=\gamma_{\rm QF}n({\bf r},t)^{3/2}$ where the effective strength $\gamma_{\rm QF}$ is defined
\begin{equation}
\gamma_{\rm QF}=\frac{32g}{3}\sqrt{\frac{a_{s}^3}{\pi}}\bigg(1+\frac{3}{2}\varepsilon_{\rm dd}^{2}\bigg).
\end{equation}
In this work we consider a quasi-one-dimensional dipolar condensate such that the transversal dynamics of the atomic cloud are effectively frozen \cite{parker_2008}. Then, a good ansatz for the wave function is $\Psi({\bf r},t)=(a_{\rho}\sqrt{\pi})^{-1}\exp(-\rho^2/2a_{\rho}^{2})\psi(x,t)$ where $a_{\rho}=\sqrt{\hbar/m\omega_{\rho}}$ defines the transverse harmonic length scale. The dimensional reduction is performed by inserting the ansatz for $\Psi({\bf r},t)$ into Eq.~\eqref{eqn:dgpe} and integrating over the transverse area of the atomic cloud. After dropping trivial energy offsets, this yields the quasi-one-dimensional generalized Gross-Pitaevskii equation
\begin{equation}\label{eqn:dgpe1d}
i\hbar\frac{\partial\psi}{\partial t}=\bigg[\frac{{\hat{p}}^{2}_{x}}{2m}+\frac{g}{2\pi a_{\rho}^2}|\psi|^2+\Phi_{\rm dd}^{\rm 1D}+\frac{2\gamma_{\rm QF}}{5\pi^{3/2}a_{\rho}^{3}}|\psi|^3\bigg]\psi.
\end{equation}
By writing Eq.~\eqref{eqn:dgpe1d} a number of approximations have been used. For the gas to be in the quasi-one-dimensional limit we require $a_\rho/\xi<1$ \cite{gorlitz_2001}, (here $\xi$ is the appropriate healing length, see Sec.~\ref{sec:hom} for further details) while the beyond mean field treatment requires $a_s n\gtrsim 0.6$ \cite{edler_2017}, and the expansion of Eq.~\eqref{eqn:int2} requires that $|\varepsilon_{\rm dd}|\sim1$. The parameters we use in our model are similar to those used in two recent dipolar experiments \cite{bottcher_2019,tanzi_2019,chomaz_2019}, in these works the trapping geometry gives a ratio of length scales $a_z/a_x=\sqrt{\omega_x/\omega_z}\simeq 0.47$, while the strength of the dipolar interaction for the atomic species $^{162}$Dy takes a typical value of $\varepsilon_{\rm dd}\sim 1.2$, similar to the values we consider through the course of this work.
The dimensionally reduced dipolar interaction appears as $\Phi_{\rm dd}^{\rm 1D}[\Psi]=\int \text{d}x'U_{\rm dd}^{\rm 1D}(x-x')|\psi(x')|^2$, where the real-space form of $U_{\rm dd}^{\rm 1D}$ is given by $U_{\rm dd}^{\rm 1D}(x)=U_{0}\mathcal{U}(|x|/a_\rho)$ with $U_0=C_{\rm dd}(1+3\cos2\alpha)/(32\pi a_{\rho}^{3})$ and $\mathcal{U}(u)=[2u-\sqrt{2\pi}(1+u^2)e^{u^2/2}\text{erfc}(u/\sqrt{2})]+(8/3)\delta(u)$ \cite{deuretzbacher_2010,sinha_2007}. Other works have even considered a general three-dimensional dipole polarization \cite{wunsch_2011}. The form of the quantum fluctuation term appearing in Eq.~\eqref{eqn:dgpe1d} is consistent with the derivation and analysis presented in Ref.~\cite{edler_2017}, who explored how the beyond mean-field term changes as the density $a_s n$ is changed. An attractive regime $-n_{\rm LHY}$ was found at low densities, while for ${n_0a_s\gtrsim0.6}$ the repulsive $n_{\rm LHY}^{3/2}$ term is recovered. In our work, the typical value of ${n_0a_s\sim10}$ is associated with the results presented throughout our work.
\begin{figure*}[t]
\includegraphics[width=1.0\textwidth]{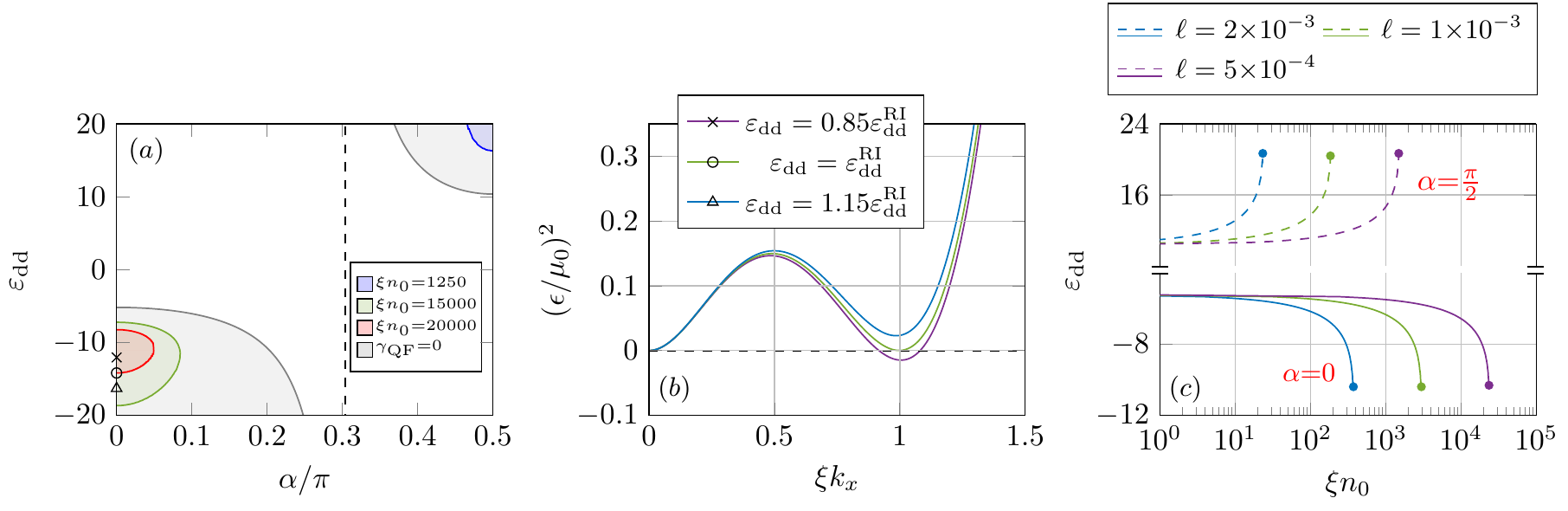}
\caption{\label{fig:rstab} Beyond-mean-field roton analysis. Panel (a) shows the roton unstable regimes in the $(\varepsilon_{\rm dd},\alpha)$ parameter space, with $\ell=5\times 10^{-4}$. Colored regions indicate roton unstable regimes for fixed $\xi n_0$=\{1250,15000,20000\} (blue, green, red respectively) obtained from Eq.~\eqref{eqn:kc} and $\eqref{eqn:ez}$. Panel (b) shows example dispersions plotted from Eq.\eqref{eqn:bdg1d} corresponding to the points indicated by the cross, circle and triangle on panel (a). Panel (c) shows how the critical roton $\varepsilon_{\rm dd}$ changes with density $\xi n_0$ for $\ell=\{5,10,20\}\times 10^{-4}$.}
\end{figure*}
To linearize Eq.~\eqref{eqn:dgpe1d} around the stationary state $\psi_0(x)$ we introduce the ansatz $\psi(x,t)=[\psi_0(x)+\delta\psi(x,t)]\exp(-i\mu t/\hbar)$ where
\begin{equation}\label{eqn:bdga}
\delta\psi(x,t)=[u(x)+v(x)]e^{-i\omega t}-[u(x)-v(x)]^{*}e^{i\omega t}
\end{equation}
describes the small-amplitude fluctuations of $\psi(x)$ and $u(x)$ and $v(x)$ represent the mode functions, $\omega$ is the associated excitation frequency and $\mu$ is the quasi-one-dimensional chemical potential. Then, by inserting the expansion of $\psi(x,t)$, including Eq.~\eqref{eqn:bdga} into Eq.~\eqref{eqn:dgpe1d} and assuming the ground state $\psi_0$ is real, we obtain the coupled equations
\begin{equation}\label{eqn:bdg}
\left[\begin{array}{cc}0 & \mathcal{H}^{\rm GP}_{\rm 1D}{-}\mu+2\mathcal{M} \\ \mathcal{H}^{\rm GP}_{\rm 1D}{-}\mu & 0\end{array}\right]\left[\begin{array}{c}u(x) \\v(x)\end{array}\right]{=}\hbar\omega\left[\begin{array}{c}u(x) \\v(x)\end{array}\right]
\end{equation}
where 
\begin{equation}
\mathcal{H}_{\rm 1D}^{\rm GP}=\frac{\hat{p}_{x}^{2}}{2m}+\frac{g}{2\pi a_{\rho}^{2}}|\psi_0|^2+\Phi_{\rm dd}^{\rm 1D}+\frac{2\gamma_{\rm QF}}{5\pi^{3/2}a_{\rho}^{3}}|\psi_0|^3
\end{equation}
defines the quasi-one-dimensional Hamiltonian appearing in Eq.~\eqref{eqn:dgpe1d}, while
\begin{equation}
\mathcal{M}=\frac{g}{2\pi a_{\rho}^{2}}|\psi_0|^2+\frac{3\gamma_{\rm QF}}{5\pi^{3/2}a_{\rho}^{3}}|\psi_0|^3+\chi
\end{equation}
gives the exchange operator. The additional nonlocal operator is defined as $\chi[f(x)]\equiv\psi_0(x)\int \text{d}x'U_{\rm dd}^{\rm 1D}(x-x')\psi_{0}(x')f(x')$. Then, the pair of Bogoliubov-de Gennes equations given by Eq.~\eqref{eqn:bdg} can be straight-forwardly decoupled. We focus on solutions for the $u(x)$ mode function, which obeys
\begin{equation}\label{eqn:u}
\bigg[\mathcal{H}_{\rm 1D}^{\rm GP}-\mu+2\mathcal{M}\bigg]\bigg[\mathcal{H}_{\rm 1D}^{\rm GP}-\mu\bigg]u(x)=(\hbar\omega)^2u(x).
\end{equation}
A similar expression for the $v(x)$ mode function can be obtained except with the bracketed terms switched in Eq.~\eqref{eqn:u}. Further details concerning the Bogoliubov-de Gennes equations are given in Appendix \ref{app:bdg}. 
\section{\label{sec:hom}Homogeneous Analysis}
\subsection{Roton Analysis}
In general the eigenvalues of the Bogoluibov de-Gennes equation Eq.~\eqref{eqn:u} must be obtained numerically. However in the homogeneous (infinite) limit one can obtain the spectrum analytically from Eq.~\eqref{eqn:u}, since the nonlocal operator $\chi$ reduces to the one-dimensional Fourier transform of the dipolar interaction. In this limit we obtain
\begin{equation}\label{eqn:bdg1d}
\epsilon^{2}{=}\epsilon_{k}^{2}{+}2n_{0}\epsilon_{k}\bigg[4a_{\rho}U_{0}\mathcal{V}_{\rm 1D}\bigg(\frac{k_{x}^2a_{\rho}^{2}}{2}\bigg){+}\frac{g}{2\pi a_{\rho}^2}{+}\frac{3\gamma_{\rm QF}\sqrt{n_0}}{5\pi^{3/2}a_{\rho}^{3}}\bigg]
\end{equation}
where the excitation energy is $\epsilon=\hbar\omega$ and the single-particle energy appearing in Eq.~\eqref{eqn:bdg1d} is $\epsilon_{k}=\hbar^2k_{x}^{2}/2m$. The term describing the Fourier transform of the dipolar interaction is defined as $\mathcal{V}_{\rm 1D}(u)=ue^{u}E_{1}(u)-1/3$ where $E_1(u)=\int_{u}^{\infty} \text{d}t\ t^{-1}e^{-t}$ defines the exponential integral. Accompanying the Bogoluibov de-Gennes energy is the chemical potential of the homogeneous system, given by
\begin{equation}\label{eqn:mu0}
\mu_0[n_0]=\frac{n_0g}{2\pi a_{\rho}^{2}}+\Phi_0+\frac{2}{5\pi^{3/2}a_{\rho}^{3}}\gamma_{\rm QF}n_{0}^{3/2},
\end{equation}
where the homogeneous dipolar potential is defined as $\Phi_0=-\varepsilon_{\rm dd}gn_0[1+3\cos 2\alpha]/8\pi a_{\rho}^{2}$. In what follows we express dimensions in terms of the natural units of the homogeneous system:  the unit of energy is the chemical potential $\mu_0$, the unit of length is the healing length $\xi=\hbar/\sqrt{m|\mu_0|}$, and the unit of time is $\hbar/|\mu_0|$. Note that as $\mu_0\rightarrow0$ the length scale $\xi\rightarrow\infty$, an inherent pathology of this choice of units. Obviously we cannot (and should not) simulate this point of the parameter space, but as we shall see we can instead simulate small $|\mu_0|$ close to the transition to the droplet state, obtaining physically sensible results.

We use the excitation spectrum defined by Eq.\eqref{eqn:bdg1d} along with the definition of the chemical potential, Eq.\eqref{eqn:mu0} to gain an understanding of the properties of the dipolar condensate in the beyond-mean-field regime. In the absence of the LHY correction, the excitation energies defined by Eq.~\eqref{eqn:bdg1d} exhibit different regimes of physical behaviour depending on the choice of parameters. In the limit of small $k_x$, the dispersion $\omega_k=k_x c_s$ is phonon-like, depending linearly on momentum such that
\begin{equation}\label{eqn:wk}
\omega_k \simeq k_x\bigg\{-\frac{4a_\rho n_0U_0}{3}+\frac{n_0g}{2\pi a_{\rho}^{2}}+\frac{3}{5\pi^{3/2}a_{\rho}^{3}}\gamma_{\rm QF}n_{0}^{3/2}\bigg\}.
\end{equation}
Here the LHY term contributes an additional density dependence to the speed of sound $c_s$ in Eq. \eqref{eqn:wk}, giving an increased value of $c_s$ in the high density phase. The point at which the roton minima touches the $k_x=0$ axis can be calculated in the following manner. First, the (squared) dispersion relation Eq.~\eqref{eqn:bdg1d} is differentiated with respect to $k_x$ and set equal to zero such that $\partial\epsilon^2/\partial k_x=0$ to obtain the two family of extrema from the dispersion, the maxon and the roton. Since we are interested in the roton, we can remove the maxon by combining this expression with the value of the dispersion set equal to zero, yielding a quartic equation in the square of the momenta $k_{x}^{2}$. This can be solved analytically to obtain 
\begin{equation}\label{eqn:kc}
\frac{k_{c}^{2}}{2}{=}{-}\frac{\mathcal{B}}{\xi^2}\bigg\{\bigg[1{+}\frac{2\mathcal{A}}{3\mathcal{B}}\bigg]{-}\sqrt{\bigg[1{+}\frac{2\mathcal{A}}{3\mathcal{B}}\bigg]^2{-}\frac{2}{\mathcal{B}\sigma^2}\bigg[1{-}\frac{\mathcal{A}}{3\mathcal{B}}\bigg]}\bigg\}
\end{equation}
which is solved simultaneously with
\begin{equation}\label{eqn:ez}
\frac{k_{c}^{2}}{2}+\frac{2n_0}{\xi}\bigg[\mathcal{A}\mathcal{V}_{\rm 1D}\bigg(\frac{k_{c}^2a_{\rho}^{2}}{2}\bigg){+}\mathcal{B}\bigg]=0.
\end{equation}
The two functions $\mathcal{A}$ and $\mathcal{B}$ that carry the dependence of the physical parameters in the problem appearing in Eqs.~\eqref{eqn:kc} and \eqref{eqn:ez} are defined as
\begin{subequations}
\begin{align}
&\mathcal{A}\equiv \frac{m\xi}{\hbar^2}4a_{\rho}U_0,\\
&\mathcal{B}\equiv \frac{m\xi}{\hbar^2}\bigg(\frac{g}{2\pi a_{\rho}^{2}}+\frac{3}{5\pi^{3/2}a_{\rho}^{3}}\gamma_{\rm QF}\sqrt{n_0}\bigg).
\end{align}
\end{subequations}
Then for a given set of physical parameters we can compute the solutions to Eqs. \eqref{eqn:kc} and \eqref{eqn:ez} numerically using an iterative procedure. Additionally, two other parameters emerge from the dimensionless analysis, the ratio of the transverse harmonic length $a_{\rho}$ and the healing length $\xi$, defined as $\sigma=a_{\rho}/\xi$. The second is the ratio of the scattering length $a_s$ and harmonic lengths $a_{\rho}$ which arrises from the LHY term, and is defined as $\ell=a_s/a_\rho$. This second dimensionless parameter has a typical value of $\ell\simeq 10^{-3}$ for dipolar gases, where the scattering length $a_s\simeq 100a_0$ and a typical radial harmonic length is $a_\rho\simeq 1\mu m$. In a previous work \cite{edmonds_2016} it was shown that for the mean-field case ($\gamma_{\rm QF}=0$) the roton instability will only appear for $a_s>0$ when $\sigma\gtrsim 0.8$, hence in what follows we assume the arbitrary value $\sigma=1$. Alternatively to our approach one can also assess the quasi-one-dimensional character of the condensate by individually considering the healing lengths associated with the contact and dipolar interactions, $\xi_{\rm CI}$ and $\xi_{\rm D}$ respectively. If one ignores the dependence of $n_0$ on $\xi$, and assumes an arbitrary droplet size (e.g. ${\sim}1\mu m$), then there will be an inconsistency between the size of the droplet and the resulting length scale $\xi$. As we will see in Sec.~\ref{sec:num}, the typical size of the droplet $L_{\rm d}(\xi)=\lambda\xi$ where $\lambda{\sim}50$ in the droplet phase. Then combining the definition of the dimensionless density $n_0=\tilde{n}_0/L_{\rm d}(\xi)$ with that of the healing lengths $\xi_j=\hbar/\sqrt{m|\mu_j|}$, one obtains the definitions 
\begin{subequations}
\begin{align}
\xi_{\rm CI}&=\frac{\lambda a_{\rho}^{2}}{2a_s\tilde{n}_0},\\
\xi_{\rm D}&=\frac{2\lambda a_{\rho}^{2}}{|\varepsilon_{\rm dd}|a_s\tilde{n}_0[1+3\cos2\alpha]},\\
\xi&=\frac{\lambda a_{\rho}^{2}}{a_s\tilde{n}_0[1-\frac{1}{4}\varepsilon_{\rm dd}(1+3\cos2\alpha)]}.
\end{align}   
\end{subequations}
With a flexible choice of physical parameters afforded by the cold atom toolbox, one can show that the ratio of length scales are $a_\rho/\xi_{\rm CI}{\sim}0.1$, $a_\rho/\xi_{\rm D}{\sim}0.2$ and $a_\rho/\xi{\sim}0.3$ where the atomic mass and dipolar strength appropriate for $^{162}$Dy has been used. One can also consider the the ratio of energies $g_{\rm 1D}n_{0}^{j}/\hbar\omega_\rho\ll1$, in which case one finds that $g_{\rm 1D}n_{0}/\hbar\omega_\rho{\sim}0.5$ and $5\times 10^{-3}$ for $a_\rho=10\mu m, a_s=10a_0$ and $a_\rho=100\mu m, a_s=10a_s$ respectively. These considerations support our claim that this model and the results obtained from it are capable of capturing the quasi-one-dimensional limit. 

Figure \ref{fig:rstab} explores the behaviour of the roton instability in the beyond-mean-field regime. Panel (a) shows the numerical solutions obtained from Eqs. \eqref{eqn:kc} and \eqref{eqn:ez} in the $(\varepsilon_{\rm dd},\alpha)$ parameter space, and we take $\ell=5\times 10^{-4}$. For $\gamma_{\rm QF}\neq 0$, two roton instabilities are observed in this system, due to the underlying quadratic dependence on $\varepsilon_{\rm dd}$ of the dispersion relation, Eq.~\eqref{eqn:bdg1d}.
\begin{figure}[t]
\includegraphics[width=\columnwidth]{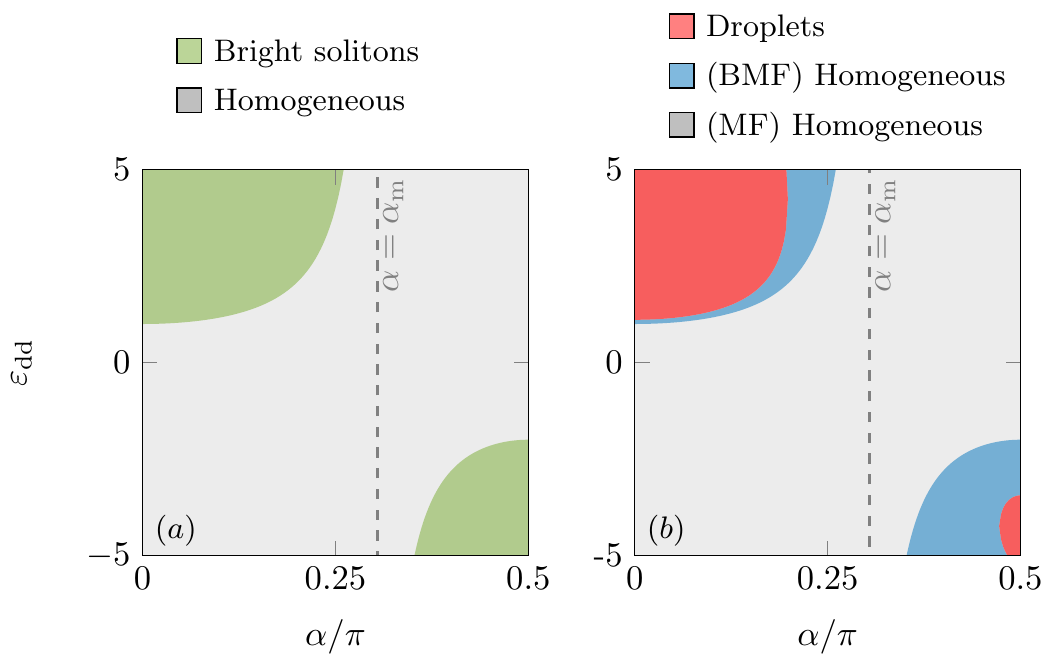}
\caption{\label{fig:phonon} Droplet phase diagram. Panel (a) shows the stable regions in the $(\varepsilon_{\rm dd},\alpha)$ parameter space in the limit $\gamma_{\rm QF}=0$. Panel (b) shows the regions where droplets are expected for $\xi n_0=10^{6}$, $\ell=10^{-3}$ and $\sigma=0.2$. The dashed lines in both panels indicate the position of the magic angle $\alpha=\alpha_m$.}
\end{figure}
\begin{figure*}[!t]
\includegraphics[width=0.9\textwidth]{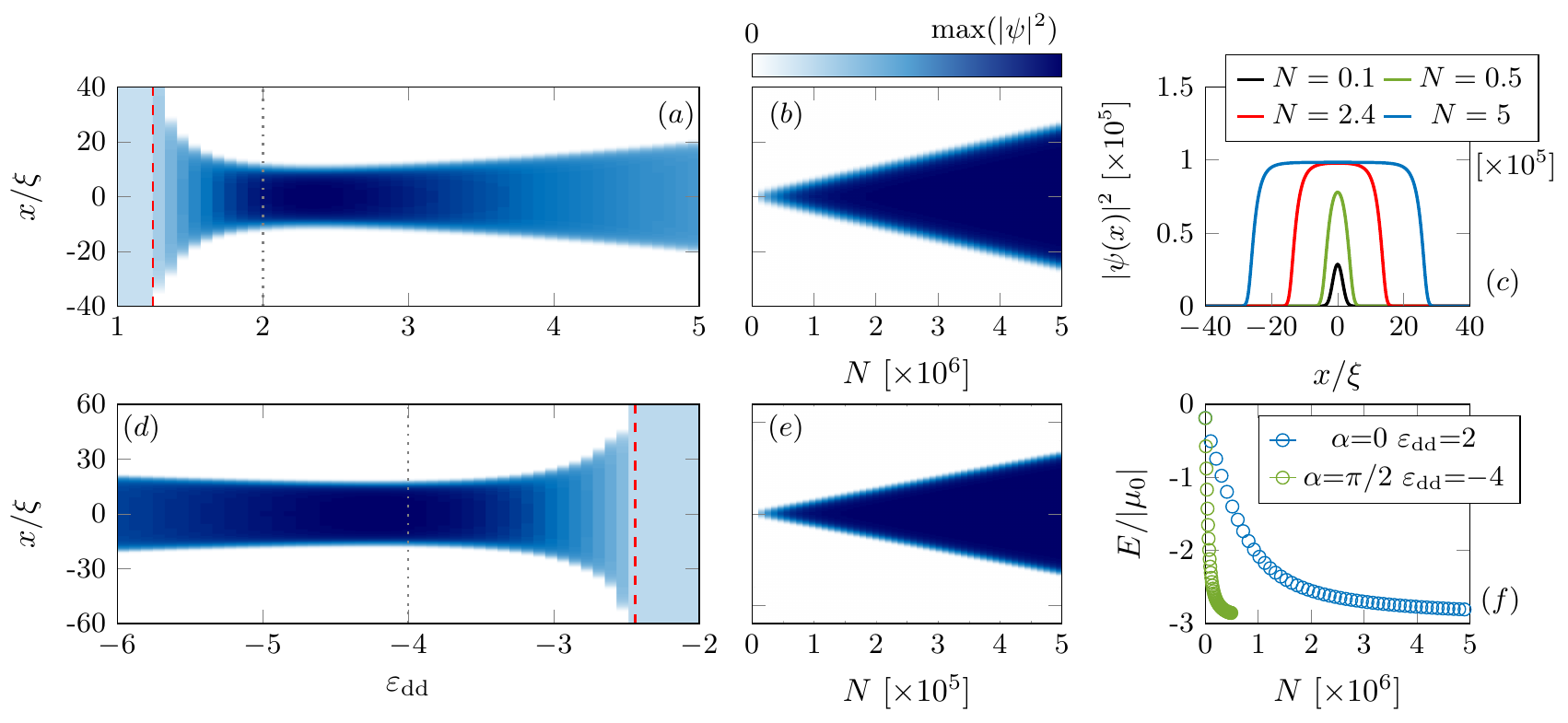}
\caption{\label{fig:gnd} Dipolar droplet ground state computation. Panels (a) and (b) show droplet ground states for fixed atom number $N=2\times 10^6$ (a) and fixed dipolar strength $\varepsilon_{\rm dd}=2$ (b), with $\alpha=0$ in both cases. Likewise panels (d) and (e) correspond to $N=2.5\times 10^5$ and $\varepsilon_{\rm dd}=-4$ with $\alpha=\pi/2$ in both cases. Panels (c) and (f) show example ground states and energies.}
\end{figure*}
Each shaded area corresponds to a different atomic density, with increasing density causing the unstable region to shrink in area. In this way the LHY term acts to stabilize the parameter space as the density is increased. The gray shaded area corresponds to the (single-valued) mean-field result in the limit $\gamma_{\rm QF}=0$. The roton unstable regions also appear for $\alpha=\pi/2$, although these are smaller in area than their $\alpha=0$ counterparts, which is attributed to the dipoles being in a side-by-side repulsive alignment, which additionally reduces the roton unstable region of the parameter space. It is an open question as to whether such a `reentrant' roton behaviour could be experimentally observed. Example dispersion relations plotted using Eq.~\eqref{eqn:bdg1d} are depicted in Fig.~\ref{fig:rstab} (b). Individual curves correspond to the cross, circle and triangle markers showing the dispersion slightly below (cross), above (triangle) and at (circle) the roton instability respectively. In these examples the density $\xi n_0=2\times 10^{4}$, while the dipole polarization $\alpha=0$. The last panel (c) of Fig.~\ref{fig:rstab} plots semi-logarithmically the value of the dipolar interaction strength $\varepsilon_{\rm dd}$ against the density $\xi n_0$ at which the roton manifests for several examples of fixed $\ell$. Each curve terminates at a maximum value of $n_0$, due to the additional repulsive LHY term overwhelming the attractive part of the dipolar interaction. Solutions for both $\alpha=0$ and $\pi/2$ are displayed, with the $\alpha=\pi/2$ solutions shifted to lower $n_0$; attributed again to the repulsive nature of this side-by-side polarization. The presence of multiple roton instabilities enriches the prospects for supersolidity within this system, which is a necessary condition for such a state to occur.   

\subsection{Quantum Depletion}
Even at zero-temperature, it is well established that some of the atoms will not be in the zero-momentum condensate mode due to interactions. This is caused by the interactions effectively mixing atoms in different finite momentum states, which is known as quantum depletion. This effect is of particular relevance for quantum dipolar droplets; since it is known to qualitatively scale with the effective diluteness parameter $n a_{s}^3$. We can compute expressions for this quantity by considering the number operator associated with the many-body Hamiltonian Eq.~\eqref{app:hmb}
\begin{equation}\label{eqn:num}
\hat{N}=N_0 + \sum_{{\bf k}\neq 0}v_{p}^{2}+\sum_{{\bf k}\neq 0}(u_{k}^{2}+v_{k}^{2})\hat{a}^{\dagger}_{\bf k}\hat{a}_{\bf k},
\end{equation}
here the $u_k$ and $v_k$ represent the momentum-space form of the \mje{three-dimensional} amplitude functions. \mje{Note that a fully-consistent calculation of the quantum depletion in the quasi-one-dimensional limit would require diagonalization of the three-dimensional Bogoliubov-de Gennes equations to obtain the mode functions $v_k$, however we can still gain some intuitive insight by considering the free-space form of the depletion.} Then the quantum depletion can be calculated in three dimensions from the second term of Eq.~\eqref{eqn:num} by taking the continuum limit such that
\begin{equation}\label{eqn:nex}
n_{\rm ex}=\int \frac{d^3 {\bf p}}{(2\pi\hbar)^3}\left[\frac{\epsilon_{\bf p}^{0}+n_{0}^{\rm 3D}U_{\Sigma}^{\bf k}}{\sqrt{{\epsilon_{\bf p}^{0}}^2+2\epsilon_{\bf p}^{0}n_{0}^{\rm 3D}U_{\Sigma}^{\bf k}}}-1\right],
\end{equation} 
here $n_{0}^{\rm 3D}$ is the homogeneous density in three-dimensions, while $U_{\Sigma}^{\bf k}=g+U_{\rm dd}({\bf k})$ is the Fourier transform of the total real-space interaction pseudo-potential, and $U_{\rm dd}({\bf k})$ is the Fourier transform of the dipole-dipole interaction, Eq.~\eqref{eqn:udk}. This quantity can be evaluated exactly for the case of $\alpha=0$ by switching to dimensionless coordinates and working in spherical polar coordinates. This yields the expression
\begin{widetext}
\begin{equation}\label{eqn:qd}
\frac{n_{\rm ex}}{n_{0}^{\rm 3D}}=\sqrt{\frac{n_{0}^{\rm 3D}a_{s}^3}{9\pi}}\left[(5\pm|\varepsilon_{\rm dd}|)\sqrt{1\pm2|\varepsilon_{\rm dd}|}+\sqrt{\frac{3}{|\varepsilon_{\rm dd}|}}(1\mp|\varepsilon_{\rm dd}|)^2\times
\begin{cases}
\text{sinh}^{-1}\bigg(\sqrt{\frac{3|\varepsilon_{\rm dd}|}{1-|\varepsilon_{\rm dd}|}}\bigg)\ \text{sgn}(\varepsilon_{\rm dd})>0,\\
\text{sin}^{-1}\bigg(\sqrt{\frac{3|\varepsilon_{\rm dd}|}{1+|\varepsilon_{\rm dd}|}}\bigg)\ \text{sgn}(\varepsilon_{\rm dd})<0.
\end{cases}\right]
\end{equation}
\end{widetext}   
in evaluating Eq.~\eqref{eqn:nex} one has to consider the sign of $\varepsilon_{\rm dd}$ as this leads to different expressions for the dipolar depletion $n_{\rm ex}$. The expression given by Eq.~\eqref{eqn:qd} scales with the diluteness parameter, which for the parameter regime studied in this work is of the order $n_{0}^{\rm 3D}a_{s}^{3}\sim3\times 10^{-6}$, and hence will not cause a significant atom loss. It is also worth commenting at this point as to the meaning of $\psi$ in this context with respect to the experiments on Helium droplets. In the context of dilute atomic gases, the quantum depletion is a small effect and hence the assumption that the system obeys a single mode equation is well supported. Experiments with Helium droplets on the other hand represent strongly correlated systems -- which do not necessarily justify such an assumption. \mje{We note that the authors of \cite{lima_2011} have also computed the dipolar depletion.}  

\subsection{Droplet Phases}
The existence of the droplet phases in the $(\varepsilon_{\rm dd},\alpha)$ parameter space depends on the balance between attractive and repulsive forces in the system. We can understand this by examining Eq.~\eqref{eqn:mu0}, the homogeneous chemical potential. In the limit $n_{0}^{\rm 3D}a_{s}^3\ll 1$ the boundary between the repulsive and attractive regions of the parameter space are independent of the density when $\mu_0=0$, which gives $\varepsilon_{\rm dd}=4/[1+3\cos(2\alpha)]$. For $n_{0}^{\rm 3D}a_{s}^{3}\sim1$ one can instead obtain two beyond-mean-field solutions to $\mu[n_0]=0$ as
\begin{equation}\label{eqn:bmf}
\varepsilon_{\rm dd}^{\pm}=\frac{\mathcal{C}_{1}\pm\sqrt{\mathcal{C}_{1}^{2}-6\mathcal{C}_0(\mathcal{C}_0-1)}}{3(\mathcal{C}_0-1)},
\end{equation}
where we have defined $\mathcal{C}_0=(128/15\pi)\ell^{3/2}\sqrt{\sigma}\sqrt{\xi n_0}+1$ and $\mathcal{C}_1=[1+3\cos(2\alpha)]/4$.
Figure \ref{fig:phonon} explores the droplets existence in the beyond-mean-field regime. Panel (a) shows the repulsive ($\mu_0>0$) and attractive ($\mu_0<0$) shaded regions, which lead to homogeneous and bright solitons respectively in the quasi one-dimensional setting. Panel (b) shows in the beyond-mean-field regime (here $n_{0}^{\rm 3D} a_{s}^{3}{\sim}3\times 10^{-6}$, close to values obtained from the experiment of Ref.~\cite{barbut_2016} which gave an approximate boundary for  dipolar droplet of $10^{-5}\lesssim n_{\rm 3D}a_{s}^3\lesssim 10^{-2}$) which for repulsive interactions gives a homogeneous ground state. However, when the mean-field phonon instability is crossed, the system remains homogeneous (blue regions). Droplets are found beyond a critical value of $\varepsilon_{\rm dd}$ obtained from Eq.~\eqref{eqn:bmf}. For $\alpha<\alpha_m$ the droplet region is larger than for $\alpha>\alpha_m$, attributed to the head-to-tail (attractive) polarization. Increasing the atomic density $n_0$ has the effect of enlarging (shrinking) these regions for $\alpha<\alpha_m$ ($\alpha>\alpha_m$). There is no roton instability in this analysis due to the choice of $\sigma=0.2$. In realizing the droplet states the sign of $C_{\rm dd}$ and $a_s$ can't be chosen arbitrarily, since the overall interactions should be attractive. Considering Fig.~\ref{fig:phonon}(a-b), the top left soliton/droplet `pockets' of both panels (a) and (b) should have $C_{\rm dd}>0$ and $a_s>0$; since $\varepsilon_{\rm dd}>1$. Likewise for the equivalent bottom right `pockets' we must instead have $C_{\rm dd}<0$ and $a_s>0$. 


\section{\label{sec:num}Numerical Simulations}
\subsection{Single Droplets}
To calculate the solutions of the generalized dipolar Gross-Pitaevskii equation Eq.~\eqref{eqn:dgpe1d} we employ a psuedo-spectral approach, the Fourier split-step method, for the results presented in this section. Due to the size of the parameter space of the extended dipolar model, which includes the strength of the dipolar interactions, the polarization angle of the dipoles, as well as the number of atoms, it is instructive to consider the ground state of Eq.~\eqref{eqn:dgpe1d} by fixing two of these physical parameters whilst varying the other. Further, in using the healing units (see Sec.~\ref{sec:hom}A ) the numerical value of the density $n_0$ appearing in the healing units is chosen such that $n_0={\rm max}(|\psi(x,t)|^2)$.

Figure \ref{fig:gnd} presents examples of the droplets spatial density $|\psi(x)|^2$ as a function of the dipolar interaction strength, $\varepsilon_{\rm dd}$ in Fig.~\ref{fig:gnd}(a) and (d) for the choices of polarization angle $\alpha=0$ and $\pi/2$ respectively. In both cases the appearance of the droplet state is not achieved for arbitrary dipole strength, but only manifests after the beyond-mean-field phonon instability is crossed, rather than the usual mean-field phonon instability corresponding to the point where the interactions become attractive in the low density limit ($n_{0}^{\rm 3D} a_{s}^3\ll 1$). In both Fig.~\ref{fig:gnd}(a) and (d) this point is indicated by a dashed red line obtained from the the solutions $\varepsilon_{\rm dd}^{\pm}$, Eq.~\eqref{eqn:bmf}. This leads to a pair of solutions for a given set of parameters, which correspond to $\alpha<\alpha_m$ and $\alpha>\alpha_m$ respectively. The width of the computed ground state is observed to be sensitive to the dipolar strength, $\varepsilon_{\rm dd}$. 
To understand the effect of changing the number of atoms in the droplet, we solve Eq.~\eqref{eqn:dgpe1d} at fixed dipolar interaction strength. Then, the dotted lines in Fig.~\ref{fig:gnd}(a) and (d) correspond to the values $\varepsilon_{\rm dd}=2$ and $\varepsilon_{\rm dd}=-4$ used in panels (b) and (e) respectively. For both polarization angles, it is found that the width of the droplet $w_{\rm drop}$ increases linearly with atom number such that $w_{\rm drop}\propto N$. The width of the droplet is largest for the $\alpha=\pi/2$ polarization angle due to the increased influence of repulsive interactions. Panel Fig.~\ref{fig:gnd}(c) shows example ground states taken from panel (b) for increasing atom number $N$. For low atom numbers (black data) the solutions resemble a bright soliton (sech-like profile) while for increasing atom number the profiles widen, developing the characteristic flat top associated with the droplet state (red and blue data). Panel (f) meanwhile shows the energy calculated from the definition
\begin{equation}\label{eqn:1den}
E=\int \text{d}x\bigg[\frac{\hat{p}^{2}_{x}}{2m}{+}\frac{g}{4\pi a_{\rho}^2}|\psi|^4{+}\frac{\Phi(x)}{2}|\psi|^2{+}\frac{4\gamma_{\rm QF}|\psi|^5}{25\pi^{3/2} a_{\rho}^{3}}\bigg]
\end{equation}
as a function of $N$ for the data presented in panel (b) and (e). The ground state energy decreases monotonically with $N$ and  $dE/dN<0$ throughout, adhering to the Vakhitov-Kolokolov stability critereon for stationary solutions of self-attractive nonlinear waves \cite{vakhitov_1973}.       

\subsection{Modulation Instability}
\begin{figure}[t]
\includegraphics[width=\columnwidth]{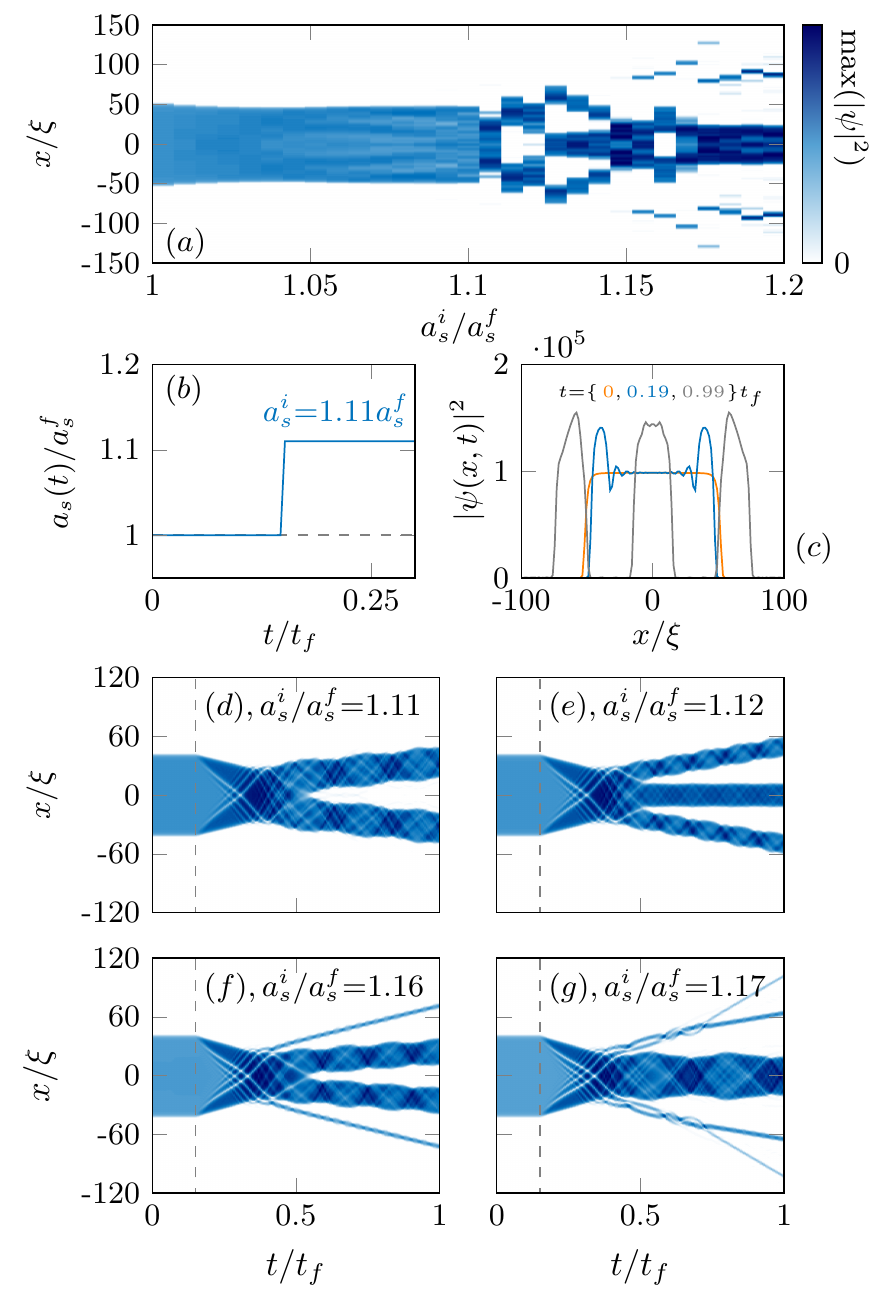}
\caption{\label{fig:mod} Modulational instability. Panel (a) shows a heatmap of the post-quench density $|\psi(x,t_f)|^2$, as a function of the quench strength $a_{s}^{i}/a_{s}^{f}$. Panel (b) shows the time-dependent scattering length $a_s(t)$ (see Eq.~\eqref{eqn:ast}), while (c) shows example quench density data for $a_{s}^{i}/a_{s}^{f}=1.12$. The three lower panels (d-g) show example dynamics for different quenches, while the dashed lines indicate the time at which the quench occurs, $t_Q=0.15t_f$.}
\end{figure}
The strength of the different nonlinearities directly determines the nature of the state that the dipolar condensate is in. By changing the $a_s$ scattering length from an initial value of $a_{s}^{i}$ to a final value $a_{s}^{f}$ such that $a_{s}^{f}<a_{s}^{i}$, a modulational instability can be induced. The instability originates from long-wavelength perturbations that cause the break-up of a waveform into pulses; coming from the growth of nonlinear excitations in the system. This effect has been used previously to investigate experimentally the stability of individual bright solitons \cite{everitt_2017,nguyen_2017} as well as dipolar droplets in a trapped three-dimensional context \cite{barbut_2018}. Figure \ref{fig:mod} investigates the effect of performing an interaction quench on an initial dipolar droplet with $\varepsilon_{\rm dd}=2$, $\alpha=0$ and $N=10^7$. The scattering length takes the time-dependent form
\begin{equation}\label{eqn:ast}
a_s(t)=a_{s}^{f}+(a_{s}^{i}-a_{s}^{f})H(t-t_{Q})
\end{equation}
where $t_Q$ defines the time at which the quench is applied, while $H(t)$ is the Heaviside function. In our simulations we take $t_Q=0.15t_f$. Figure \ref{fig:mod}(a) shows the atomic density $|\psi(x,t_f)|^2$ of the dipolar gas as a function of the quench strength $a_{s}^{i}/a_{s}^{f}$ taken at the final point of the numerical integration $t=t_f$. For modest quench strengths ($a_{s}^{i}/a_{s}^{f}\lesssim 1.1$) the droplets shape remains intact, with the exception of the excitation of surface waves. Above a critical quench strength the droplet undergoes the modulation instability, and in general breaks apart into smaller droplets and dipolar bright solitons, as well as the emission of low density radiation. Panel (b) and (c) show respectively the quench protocol of Eq.~\eqref{eqn:ast} and example dynamics for $a_{s}^{i}/a_{s}^{f}=1.12$. The lower row of panels (d-g) of Fig.~\ref{fig:mod} show individual examples of the quench dynamics for increasing $a_{s}^{i}/a_{s}^{f}$. Here we observe both even and odd numbers of droplets, while panel (f) shows a situation where two droplets and two bright solitons are created after the quench. For larger quench strengths, such as that presented in panel (g), a central droplet is produced along with increasing amounts of radiation, in this example short-lived bright soliton bound states are also observed. In general we find that the onset of the modulational instability (and the final state after the quench) depend strongly on the atom number. If for example the initial number of atoms in the droplet is reduced, then the modulational instability occurs at larger values of $a_{s}^{i}/a_{s}^{f}$ compared to a larger initial atom number. The generation of multiple droplets as presented here using the modulational instability relies on being able to tune the scattering length $a_s$ of the condensate which can in turn lead to significant atomic losses. To address this, there are proposals to produce condensates with attractive interactions with reduced noise and greater control over the final experimental state of the system \cite{billam_2011,edmonds_2018}.

\subsection{Collisional Population Transfer and Droplet Fission}  
The coherent nature of the superfluid state provides a convenient tool to explore quantum mechanical phenomena at macroscopic length scales. One striking manifestation of matter-waves coherence is the so-called Josephson effect. Here, two superconductors or superfluids which are separated by an insulting barrier can experience a current, originating from atomic tunneling between the two superconductors/superfluids. The dipolar droplets represent an interesting addition to the superfluid family, since they are effectively an isolated (finite) region of homogeneous fluid, so understanding their binary dynamics is expected to yield novel phenomena.
To investigate the basic physics of binary droplet dynamics, we perform simulations with an initial state of the form
\begin{equation}\label{eqn:psi0}
\psi_0(x)=\sum_{n=\pm}\psi(x-x_n)e^{imv_n x/\hbar+i\delta_n}.
\end{equation} 
This constitutes a symmetric state comprising two droplets whose centres are initially separated by a distance $x_+-x_-=60\xi$, which are traveling towards each other at constant velocity $v_\pm=\pm v_0$. The initial phase difference $\delta=\delta_+-\delta_-$ between the two droplets is $\delta\in[0,2\pi]$.
\begin{figure}[t]
\includegraphics[width=\columnwidth]{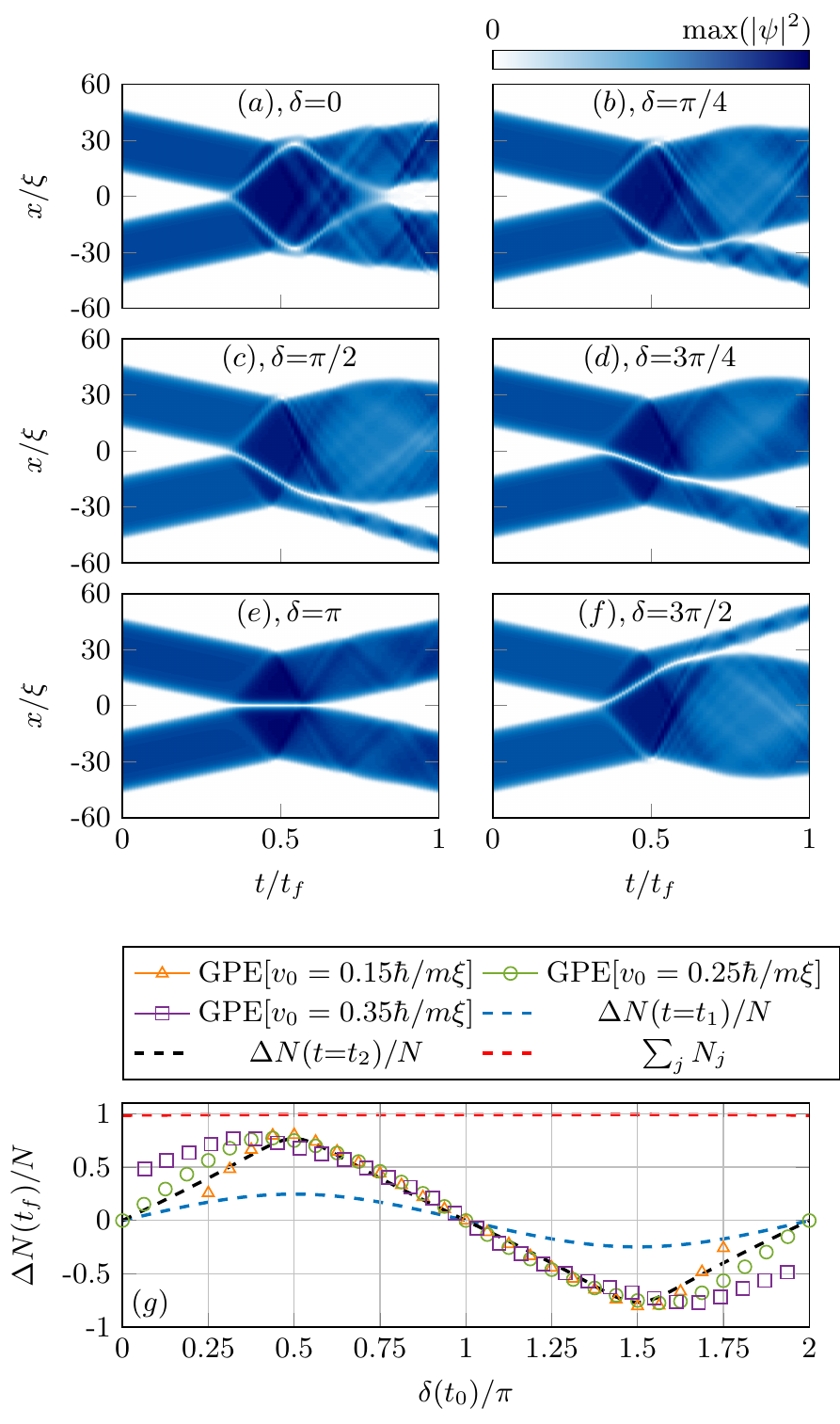}
\caption{\label{fig:jos} Population transfer. Panels (a-f) show space-time dynamics for different initial phase differences, $\delta$. The total integration time is $t_f=140\hbar/|\mu_0|$. Panel (g) shows a comparison of the final droplet populations at $t=t_f$ calculated using Eq.~\eqref{eqn:nj}.}
\end{figure}
In Figure \ref{fig:jos} we present results of droplet collisions using the initial state defined by Eq.~\eqref{eqn:psi0}. We take for the physical parameters $\varepsilon_{\rm dd}=2$, $N=3\times 10^{6}$, $\ell=10^{-3}$, $\sigma=0.2$, and the total length of each numerical integration is $t_f=140\hbar/|\mu_0|$. Figures \ref{fig:jos} (a-f) show space-time plots for different initial phase differences, $\delta$ and initial velocity $m\xi v_0/\hbar=0.25$. We observe that post collision two droplets emerge - with different populations (and sizes) that depend on the choice of initial phase difference. For example, choosing $\delta=\pi$ (panel \ref{fig:jos} (e)) produces two droplets with an equal number of atoms present in each droplet post collision. The presence of excitations in the form of sound waves can be seen here post collision, reflecting back and forth inside each droplet (viz. Eq.~\eqref{eqn:wk}), which is indicative of non-integrable dynamics. To quantify this change in population, we can calculate the population difference between the droplets at $t=t_f$ as a function of the initial phase difference $\delta(t_0)$. The population difference is defined as $\Delta N(t_f)=N_1(t_f)-N_2(t_f)$ where
\begin{equation}\label{eqn:nj}
N_j(t)=\int^{x_{j+}}_{x_{j-}}\text{d}x |\psi_j(x,t)|^2.
\end{equation}
Here each integral computes the number of atoms in the individual droplets between the edges of the droplet given by $x=x_{j\pm}$. Panel \ref{fig:jos}(g) shows some example results with different choices of initial velocity, $m\xi v_0/\hbar=0.15,0.25,0.35$ (triangle, circle and square markers). Only droplet collisions for $m\xi v_0/\hbar=0.25$ represents a situation where two droplets emerge post collision for the full range of phase differences between $\delta=0$ and $\delta=\pi$, due to the existence of bound states (droplet molecules) and droplet fission at smaller and larger initial velocities respectively, breaking droplet number conservation. As such, a parameter window exists where one can compare the collisional transfer to the Josephson effect. Then, the semi-classical Josephson equations for a superfluid are written as \cite{pethick_smith}
\begin{subequations}\label{eqn:je}
\begin{align}
\frac{d}{dt}\Delta N(t)=\frac{J}{\hbar}\sqrt{N_1(t)N_2(t)}\sin(\delta(t)),\\
\frac{d}{dt}\delta(t)=\frac{J}{\hbar}\bigg[\frac{N_1(t)}{N_2(t)}-\frac{N_2(t)}{N_1(t)}\bigg]\cos(\delta(t)).
\end{align}
\end{subequations}
Here, the parameter $J$ describes the strength of the coupling between the two superfluids. We will use the final integration time $t=t_f$ to fit our model to the numerical simulations of the extended GPE (green circles), since the total integration time determines the nature of the observed Josephson oscillations. Panel \ref{fig:jos} (g) shows a comparison between the populations of the droplets at $t=t_f$, calculated from the numerical solutions to the Josephson equations. The blue and black dashed lines are obtained from the pair of Josephson relations Eq.~\eqref{eqn:je}, for $t_f=0.5\hbar/J$ and $t_f=1.75\hbar/J$ respectively. For small $t_f$, (blue dashed) the oscillation is linear, and does not follow the extended GPE data (green circles). For larger $t_f$, (black dashed) the oscillation exhibits a stronger nonlinear character, and follows the extended GPE data quite well. It would be interesting to study this effect in more detail, to understand if this analogy with the Josephson effect can be extended, for example into the highly nonlinear regime. The related experiment of Ref.~\cite{nguyen_2014} studied the collisions of (trapped) matter-wave solitons in an attractive condensate of $^{7}$Li atoms. They investigated how the relative phase $\Delta\phi$ between the solitons affected the collision dynamics of the solitons, observing constructive and destructive matter-wave interference at the point of collision for in-phase ($\Delta\phi=0$) and out-of-phase ($\Delta\phi=\pi$) respectively. They also observed the in-phase collapse of the wave function above a critical atom number. Our work reveals similar phase-sensitive dynamics, as well as population transfer between the droplets post collision. Unlike the solitons \cite{parker_2008a} the droplets cannot undergo collapse, which is suppressed by the repulsive LHY contribution. Hence we are able to probe the dynamics for all relative phases.
\begin{figure}[t]
\includegraphics[width=\columnwidth]{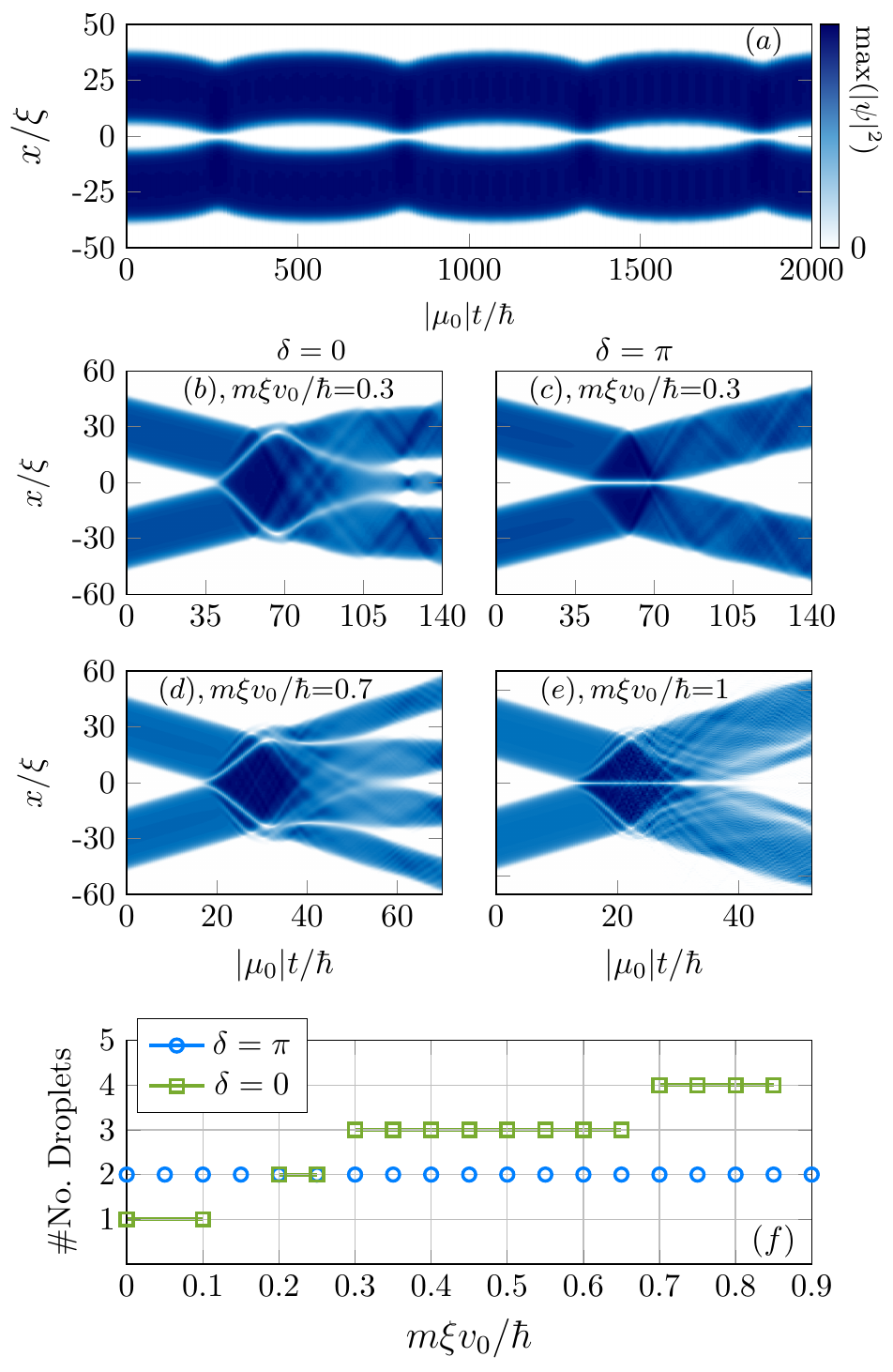}
\caption{\label{fig:coll} Droplet collisions. A long-lived droplet dimer is shown in panel (a), while panels (b-e) show in and out-of-phase dynamics for different initial velocities. (f) computes the number of droplets as a function of the initial velocity, $v_0$ for in-phase $\delta=0$ collisions.}
\end{figure}

To understand the effect of the droplets initial velocity on the dynamics, Figure \ref{fig:coll} shows simulations of droplet collisions with fixed initial phase difference, for $\delta=0,\pi$. Panel (a) shows a long-lived droplet dimer formed from two initially out-of-phase stationary droplets. Then, example dynamics are shown in panels (b-e) for different finite initial velocities (see individual captions). The in-phase collisions demonstrate that the droplets undergo fission \cite{pigier_2001,musslimani_2001,dingwall_2018} with multiple droplet states produced post-collision, shown in panels (b) and (d). For the out-of-phase collisions ($\delta=\pi$) two out-going droplets are observed, for low incoming velocity (panel (c)) where small amounts of sound are produced post collision. For a greater initial velocity (panel (e)), larger amounts of sound and radiation are produced, still with two out-going droplets. Finally panel (f) computes the number of droplets post collision for in and out-of-phase collisions, as a function of the initial velocity. For even larger velocities ($m\xi v_0/\hbar\sim 1$) the delicate balance of kinetic and potential energy that maintains the droplet is violated, and the droplet breaks apart into atomic radiation. \mje{The splitting of a single droplet into multiple smaller droplets is also noted to occur with classical fluids.} Experimental investigations have also explored the collisions of droplets in a binary system \cite{ferioli_2019}. It is interesting to compare the results of this experiment with our findings. Ferioli et al. found that the outcome of the droplets collisions depends critically on the relative velocity of the droplet pair. For smaller incident velocities the droplets could form a bound state, while at larger velocities the droplets remain separated post collision. Our results concerning dipolar droplets reveal that the droplets can instead split into increasing numbers of smaller droplets as the speed of collision increases. It should be noted that as well as the beyond-mean-field LHY term the authors of Ref.~\cite{ferioli_2019} also model three body effects, which we have not considered in our work, and would be a natural and interesting extension for a future study.

\section{\label{sec:conc}Conclusions and Outlook}
In this work we have investigated the properties of a quasi-one-dimensional dipolar Bose-Einstein condensate in the presence of quantum fluctuations. 
By calculating the excitations of this system within the framework of the Bogoliubov-de Gennes formalism, we identified regimes unstable to the roton instability. Interestingly, the roton unstable regions are found in general to appear in pairs for a given dipole polarization angle; this is due to the underlying quadratic dependence on the dipolar strength from the LHY contribution to the excitation energy.

We examined the nature of the ground states of this system, observing the appearance of droplet phases in regions of the parameter space where the total interactions are net attractive. By applying an interaction quench to a single large droplet, the nature of the modulation instability was investigated. It was found that for moderate quenches, both even and odd numbers of droplets can be generated. For larger quenches, bright solitons and increasing amounts of atomic radiation are produced, suggesting a window of parameters for useful quenches. The collisional properties of droplets were also explored. By modulating the initial phase difference between the droplets, atomic population transfer was observed between the droplets. This was interpreted and compared with the superfluid Josephson effect, finding good agreement in a window of the parameter space. Droplet fission was also observed as the initial velocity of the droplet was increased. 

It would be interesting in the future to understand how a harmonic trap changes the physics of the one-dimensional droplet, and in particular how the interplay of quenching both the interactions and trapping strength changes the number of droplets produced. One could also use this model to understand supersolid phases in the one-dimensional context, as well as studying droplets and their dynamics with models that incorporate higher-dimensional effects \cite{knight_2019,blakie_2020}. Finally, it would also be beneficial to understand the lifetime of the one-dimensional droplet, which can be computed from three-body atomic recombinations \cite{bottcher_2019a}.

\section{Acknowledgements}

We thank Jing Li for discussions and Thomas Flynn for comments on the manuscript. This work was supported by the Ministry of Education, Culture, Sports, Science (MEXT)-Supported Program for the Strategic Research Foundation at Private Universities ``Topological Science (Grant No. S1511006). NGP acknowledges support from the Engineering and Physical Sciences Research Council (Grant No. EP/M005127/1) for support. TB acknowledges support from the Engineering and Physical Sciences Research Council (Grant No. EP/R51309X/1) for support.

\appendix
\section{\label{app:qf}Quantum Fluctuations}
In this appendix we compute the form of the dipolar LHY term \cite{lima_2011,schutzhold_2006} including the effect of the polarization angle of the dipoles. The LHY correction arises from correcting the otherwise divergent ground state energy of a homogeneous gas of bosons. 
The many-body Hamiltonian $\hat{H}$ for a gas of $N$ interacting dipoles of volume $V$ can be written as \cite{ueda_book}
\begin{align}\nonumber
&\hat{H}=\frac{n_{0}^{\rm 3D} N}{2}U_{\Sigma}^{\bf k}+\sum_{\bf k\neq 0}\sqrt{\varepsilon_{\rm k}^{0}(\varepsilon_{\rm k}^{0}+2n_{0}^{\rm 3D}U_{\Sigma}^{\bf k})}\hat{a}^{\dagger}_{\rm k}\hat{a}_{\rm k} \\&{-}\frac{1}{2}\sum_{\bf k\neq 0}\bigg[n_{0}^{\rm 3D}U_{\Sigma}^{\bf k}{+}\varepsilon_{\rm k}^{0}{-}\sqrt{\varepsilon_{\rm k}^{0}(\varepsilon_{\rm k}^{0}{+}2n_{0}^{\rm 3D}U_{\Sigma}^{\bf k})}\bigg],\label{app:hmb}
\end{align}
here $\hat{a}_{\rm k}$ ($\hat{a}_{\rm k}^{\dagger}$) defines the annihilation (creation) operator for a quasi-particle with momentum ${\bf k}$ and energy $\varepsilon_{\rm k}^{0}=\hbar^2{\bf k}^2/2m$, and $U_{\Sigma}^{\bf k} = g + U_{\rm dd}({\bf k})$ is used as short-hand for the total Fourier transform of the interaction pseudo-potential, which is defined as \cite{lahaye_2009}
\begin{equation}\label{eqn:udk}
U_{\rm dd}({\bf k})=\frac{C_{\rm dd}}{3}\bigg[3\frac{(k_x\sin\alpha+k_z\cos\alpha)^2}{k_{x}^2+k_{y}^2+k_{z}^2}-1\bigg]
\end{equation}
\begin{figure}
\includegraphics[width=0.9\columnwidth]{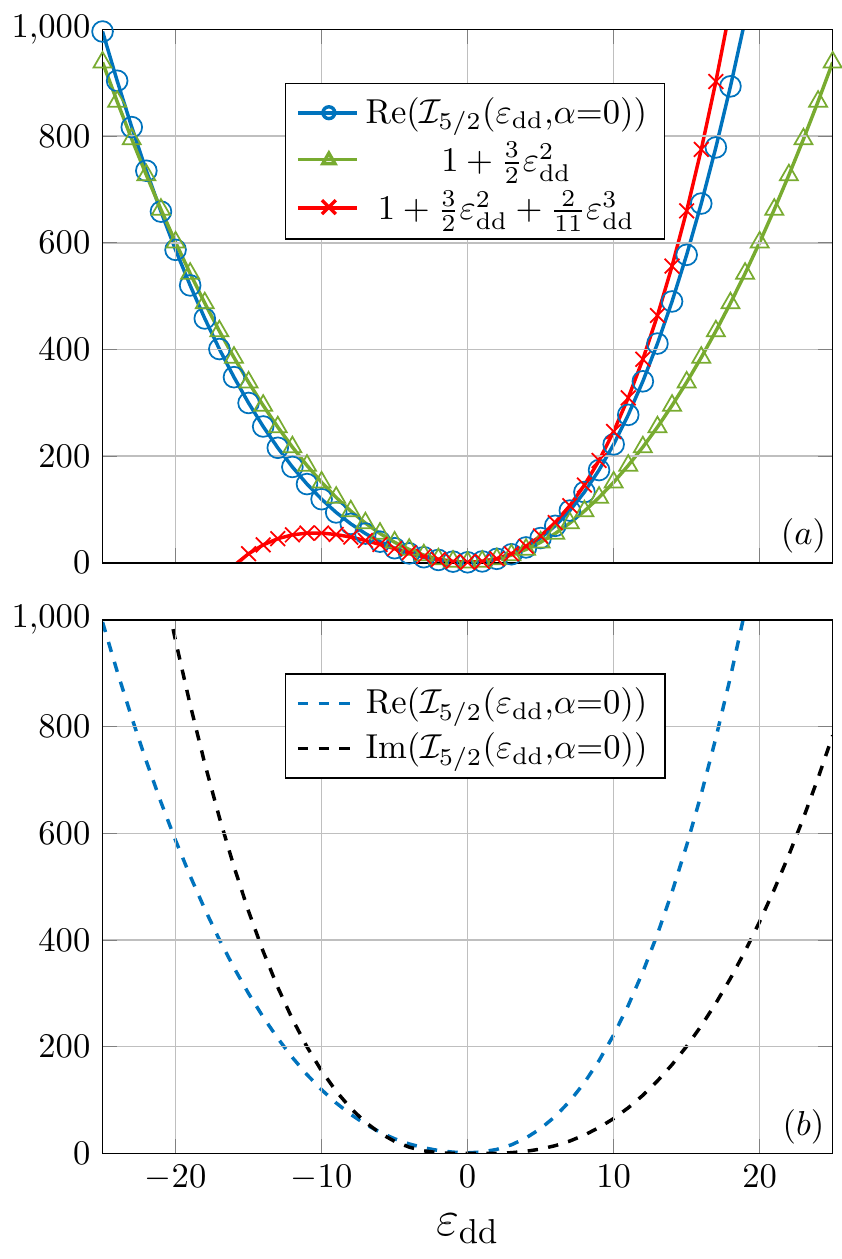}
\caption{\label{fig:int}Comparison of approximations. In (a) the blue circles show the real part of the exact (numerical) form of Eq.~\eqref{eqn:int1} for $n=5/2$, while the green triangles and red crosses show the quadratic and cubic approximations respectively of Eq.~\eqref{eqn:eqf_cub}. (b) shows the real and imaginary parts of Eq.~\eqref{eqn:int1}.}
\end{figure}
where $k_i$ defines the cartesian components of the momentum, and $\alpha$ is the dipole polarization angle in the $x$-$z$ plane (see Fig.~\ref{fig:dipole} (a)). The ground state energy is defined in the usual quantum mechanical way as $E_{\rm QF}=\langle\Psi_{\rm k\neq0}|\hat{H}|\Psi_{\rm k\neq0}\rangle$, and since $\hat{a}_{\bf k\neq 0}^{\dagger}\hat{a}_{\bf k\neq 0}|\Psi_{\rm k\neq0}\rangle=0$, the summation term on the first line of Eq.~\eqref{app:hmb} does not contribute to $E_{\rm QF}$. Converting the remaining summation on the second line of Eq.~\eqref{app:hmb} to an integral using 
\begin{equation}
\sum_{\rm{\bf k}\neq 0}\rightarrow V\int \frac{{\rm d}^3{\bf k}}{(2\pi)^3} 
\end{equation}
we find that this term diverges as ${\bf k}\rightarrow\infty$. Since the ground state energy $E_{\rm QF}$ should be a real finite quantity, we can \textit{renormalize} this term to give a finite result. Then, expanding this term in inverse power of the kinetic energy, one can show that
\begin{equation}\label{app:exp}
n_{0}^{\rm 3D}U_{\Sigma}^{\rm k}{+}\varepsilon_{\rm k}^{0}{-}\sqrt{\varepsilon_{\rm k}^{0}(\varepsilon_{\rm k}^{0}{+}2n_{0}^{\rm 3D}U_{\Sigma}^{\rm k})}\simeq {n_{0}^{\rm 3D}}^2\frac{{U_{\Sigma}^{\rm k}}^2}{2\varepsilon_{\bf k}^{0}}.
\end{equation} 
Then, a finite value for the ground state energy can be obtained by subtracting this final term in Eq.~\eqref{app:exp} from the summation on the second line of Eq.~\eqref{app:hmb}. Hence we write
\begin{align}\nonumber
\frac{E_{\rm QF}}{V}{=}&-\frac{1}{2}\int \frac{{\rm d}^3 {\bf k}}{(2\pi)^3} \bigg[n_{0}^{\rm 3D}U_{\Sigma}^{\bf k} {-} \sqrt{\varepsilon_{\rm k}^{0}(\varepsilon_{\rm k}^{0}{+}2n_{0}^{\rm 3D}U_{\Sigma}^{\rm k})} \\&{-} {n_{0}^{\rm 3D}}^2\frac{{U_{\Sigma}^{\rm k}}^2}{2\varepsilon_{\bf k}^{0}}\bigg],\label{eqn:eqf3d}
\end{align}
Although the integral defined by Eq.~\eqref{eqn:eqf3d} is convergent, it does not in general exist in closed form. After integrating out the radial momentum from Eq.~\eqref{eqn:eqf3d}, one can show that this expression can be re-written as
\begin{equation}\label{eqn:eqf_s}
\frac{E_{\rm QF}}{V}=\frac{8\sqrt{2}}{15}\frac{n_{0}^{\rm 3D}g}{(2\pi)^2}\bigg(\frac{2mn_{0}^{\rm 3D}g}{\hbar^2}\bigg)^{3/2}\mathcal{I}_{5/2}(\varepsilon_{\rm dd},\alpha)
\end{equation} 
where
\begin{equation}\label{eqn:int1}
\mathcal{I}_{n}(\varepsilon_{\rm dd},\alpha)=\int\frac{\text{d}\Omega}{4\pi}\bigg\{1+\varepsilon_{\rm dd}[3L(\theta,\phi,\alpha)^2-1]\bigg\}^{n}.
\end{equation}
Here the short-hand $L(\theta,\phi,\alpha)=\sin\theta\cos\phi\sin\alpha+\cos\theta\cos\alpha$ has been used. Although the integral defined by Eq.~\eqref{eqn:int1} does not exist in closed form, on physical grounds we are interested in situations where the parameter $\varepsilon_{\rm dd}\gtrsim1$ typically, so we expand the integrand in the first few powers of $\varepsilon_{\rm dd}$, which will allow us to identify an appropriate analytical approximation to Eq.~\eqref{eqn:eqf_s}. Then taking $n=5/2$, Eq.~\eqref{eqn:int1} becomes
\begin{align}\nonumber
&\mathcal{I}_{5/2}(\varepsilon_{\rm dd},\alpha)=\int\frac{\text{d}\Omega}{4\pi}\bigg\{1{+}\frac{5\varepsilon_{\rm dd}}{2}[3L(\theta,\phi,\alpha)^2{-}1]\\&{+}\frac{15\varepsilon_{\rm dd}^{2}}{8}[3L(\theta,\phi,\alpha)^2{-}1]^2{+}\frac{15\varepsilon_{\rm dd}^3}{72}[3L(\theta,\phi,\alpha)^2{-}1]^3{+}\dots\bigg\}.\label{eqn:int2}
\end{align}
Evaluation of the integrals appearing in Eq.\eqref{eqn:int2} can be accomplished using the result 
\begin{equation}\label{eqn:trig_int}
\int\frac{\text{d}\Omega}{4\pi}L(\theta,\phi,\alpha)^{2n}=\frac{1}{2n+1}, n\in\mathds{Z}^{\ge},
\end{equation}
then after collating the results of Eqs.~\eqref{eqn:eqf_s}, \eqref{eqn:int2} and \eqref{eqn:trig_int} we arrive at  
\begin{equation}\label{eqn:eqf_cub}
\frac{E_{\rm QF}}{V}=\frac{64}{15}g(n_{0}^{\rm 3D})^2\sqrt{\frac{n_{0}^{\rm 3D}a_{s}^{3}}{\pi}}\bigg(1+\frac{3}{2}\varepsilon_{\rm dd}^{2}+\frac{2}{11}\varepsilon_{\rm dd}^3\bigg).
\end{equation}
In Fig.~\ref{fig:int}(a) we compare the quadratic and cubic forms of Eq.~\eqref{eqn:eqf_cub} with the exact numerically calculated value of Eq.~\eqref{eqn:eqf_s}. Figure \ref{fig:int}(a) shows three sets of data. The blue circles show the exact value of Eq.~\eqref{eqn:int1} computed numerically, while terms up-to quadratic (green triangles) and cubic (red crosses) orders are plotted using Eq.~\eqref{eqn:eqf_cub}. For $\varepsilon_{\rm dd}>0$, it is clear that the cubic form of Eq.~\eqref{eqn:eqf_cub} gives closer agreement with the exact form of $\mathcal{I}_{\rm 5/2}(\varepsilon_{\rm dd},\alpha)$. However for $\varepsilon_{\rm dd}<0$ the agreement breaks down, the difference being $\sim50\%$. The quadratic approximation is not quite as good as that of the cubic one for positive $\varepsilon_{\rm dd}$, but gives closer agreement for negative $\varepsilon_{\rm dd}$, compared to the cubic form. As such, we adopt a quadratic approximation, which overall allows us to study the role of the quantum fluctuations from a numerical and analytical viewpoint, hence the quadratic nature of Eq.~\eqref{eqn:eqf} of the main text. Panel (b) shows a comparison of the real (black-dashed) and imaginary (blue-dashed) parts of $\mathcal{I}_{\rm 5/2}$. Note for $-0.5\leq\varepsilon_{\rm dd}\leq1$, one has Im$(\mathcal{I}_{\rm 5/2})=0$. For most of our results we only consider values of $|\varepsilon_{\rm dd}|\lesssim5$ (see Fig.~\ref{fig:phonon}-\ref{fig:coll}) for which the imaginary contribution to the LHY term is still relatively small compared to the real part. Then in Fig.~\ref{fig:rstab} we consider dipolar interaction strengths $|\varepsilon_{\rm dd}|{\sim }20$, which according to Fig.~\ref{fig:int}(b) would be problematic since Im$(\mathcal{I}_{\rm 5/2})\gg1$. We argue that it may still be possible to observe such an effect. In Ref.~\cite{edmonds_2016} we previously explored where the roton can appear for $\sigma\lesssim1$ if instead one has sgn$(a_s)<0$. In these regions the rotons are predicted closer to $\varepsilon_{\rm dd}=0$, exactly where the full calculation of $\mathcal{I}_{\rm 5/2}$ is purely real. This offers a possible opportunity for future studies of this interesting effect.

\section{\label{app:bdg}Bogoliubov-de Gennes Equations}
In this appendix we give further details of the derivation of the beyond-mean-field Bogoliubov-de Gennes equations of the text, Eqs.~\eqref{eqn:bdg}. The derivation with the methodology we employed to solve Eq.~\eqref{eqn:u} is qualitatively the same for either mode function, here we focus on the $u(x)$ mode function without loss of generality. To proceed, we note that the time-independent beyond-mean-field dipolar Gross-Pitaevskii equation $[\mathcal{H}_{\rm 1D}^{\rm GP}-\mu]\phi_j=\epsilon_{j}^{\rm GP}\phi_j$ (obtained from Eq.~\eqref{eqn:dgpe1d} by the substitution $\psi(x,t)=\phi_{j}(x)\exp(-i[\epsilon_{j}^{\rm GP}+\mu]t/\hbar)$) possesses a spectral basis with orthonormal modes $\phi_j(x)$ and corresponding energies $\epsilon_{j}^{\rm GP}$. Since the condensate mode has been removed, the resulting quasiparticle modes are automatically orthogonal to the condensate. Then by making the expansion
\begin{equation}\label{eqn:sp}
u(x)=\sum_{\lambda}c_{\lambda}\phi_{\lambda}(x)
\end{equation}
and inserting Eq.~\eqref{eqn:sp} into \eqref{eqn:u} whilst premultiplying by $\phi_{\gamma}^{*}(x)$ and using the orthonormal property of the spectral basis states 
\begin{equation}
\int \text{d}x \phi_{\gamma}^{*}(x)\phi_{\lambda}(x)=\delta_{\gamma\lambda}, 
\end{equation}
we obtain the matrix-valued equation for the $\omega$ eigenvalues 
\begin{equation}\label{eqn:mham}
\sum_{\lambda}\bigg[\delta_{\gamma\lambda}\epsilon_{\rm \lambda}^{\rm GP}+2\mathcal{M}_{\rm\gamma\lambda}\bigg]\epsilon_{\rm \lambda}^{\rm GP}c_{\lambda}=(\hbar\omega_{})^2c_{\gamma},
\end{equation}
with the exchange matrix elements $\mathcal{M}_{\gamma\lambda}=\langle\phi_\gamma|\mathcal{M}|\phi_\lambda\rangle$ defined as 
\begin{align}\nonumber
\mathcal{M}_{\gamma\lambda}=&\int \text{d}x\phi_{\gamma}^{*}(x)\bigg[\frac{g}{2\pi a_{\rho}^{2}}|\psi_0|^2+\frac{3\gamma_{\rm QF}}{5\pi^{3/2}a_{\rho}^{3}}|\psi_0|^3\bigg]\phi_{\lambda}(x)\\&+\frac{1}{2\pi}\int \text{d}k\varphi_{\gamma}^{*}(-k)U_{\rm dd}^{\rm 1D}(k)\varphi_{\lambda}(k),\label{eqn:mgl}
\end{align}
with $\varphi_\nu(k)=\int \text{d}x\phi_\nu(x)\psi_0(x)\exp(ikx)$ \cite{bland_thesis}.


\end{document}